\documentclass[twocolumn]{article}
\usepackage{amsmath}
\usepackage{graphicx}
\graphicspath{ {./images/} } 
\usepackage{geometry}
\usepackage{floatrow}
\usepackage{layout}
\usepackage{amssymb} 
\usepackage{multirow}
\usepackage{caption}
\usepackage{authblk}
 \usepackage{indentfirst}
\usepackage{comment}
\usepackage{abstract}
\usepackage{balance}
\usepackage{cite}

\makeatletter
\def\preparefootins{%
\global\rcol@footinsskip\skip\footins
\global\skip\footins\z@
\global\count\footins\z@
\global\dimen\footins2\textheight}
\makeatother

\usepackage{xcolor}
\definecolor{darkblue}{rgb}{0.15,0.35,0.55}
\definecolor{reddish}{rgb}{0.65, 0.2, 0.2}
\definecolor{purplish}{rgb}{.2, .2, .5}

\usepackage{color}   
\usepackage{hyperref}
\hypersetup{
    colorlinks=true, 
    linktoc=black,     
    linkcolor=blue,  
    citecolor=red
}
 
\usepackage[rightcaption]{sidecap}
\usepackage{graphicx}
\graphicspath{ {./images/} }

\newcommand{\m}{\mu}
\newcommand{\n}{\nu}
\newcommand{\p}{\vec\nabla}

\newcommand{\vev}[1]{{\left< {#1} \right>}}

\newcommand{\cur}[1]{{\left( {#1} \right)}}
\newcommand{\squ}[1]{{\left[ {#1} \right]}}

\newcommand{\cA}{{\mathcal A}}
\newcommand{\cB}{{\mathcal B}}

\newcommand{\cD}{{\mathcal D}}
\newcommand{\cK}{{\mathcal K}}
\newcommand{\cL}{{\mathcal L}}
\newcommand{\cN}{{\mathcal N}}
\newcommand{\cP}{{\mathcal P}}
\newcommand{\cQ}{{\mathcal Q}}

\newcommand{\cM}{{\mathcal M}}

\newcommand{\cE}{{\mathcal E}}

\newcommand{\cR}{{\mathcal R}}

\newcommand{\cJ}{{\mathcal J}}

\usepackage{tocbasic}[2016/05/10]
\DeclareTOCStyleEntry[indent=0em,numwidth=3em]{tocline}{section}
\DeclareTOCStyleEntry[indent=1.5em,numwidth=3.5em]{tocline}{subsection}

\usepackage{sectsty}
\sectionfont{\centering\fontsize{10}{10}\selectfont \MakeUppercase}
\subsectionfont{\centering\fontsize{10}{10}\selectfont}
\subsubsectionfont{\bf\centering\fontsize{10}{10}\selectfont}

\begin{document}
\title{\textbf{\Large{Dynamical chemistry:~non-equilibrium \\ effective actions for reactive fluids}}}
\author{{ \normalsize Michael J. Landry}}
\affil{{\it\normalsize Department of Physics, Center for Theoretical Physics, \\ \it Columbia University, 538W 120th Street, New York, NY, 10027, USA}}
\date{}
\twocolumn[
 \maketitle
 \begin{onecolabstract} 
    We present two approaches for describing chemical reactions taking place in  fluid phase.  The first method mirrors the usual derivation of the hydrodynamic equations of motion by relating conserved---or to account for chemical reactions, non-conserved---currents to local-equilibrium parameters. The second method involves a higher-brow approach in which we attack the same problem from the perspective of non-equilibrium effective field theory (EFT). Non-equilibrium effective actions are defined using the in-in formalism on the Schwinger-Keldysh contour and are therefore capable of describing thermal fluctuations and dissipation as well as quantum effects. The non-equilibrium EFT approach is especially powerful as all terms in the action are fully specified by the symmetries of the system; in particular the second law of thermodynamics does not need to be included by hand, but is instead derived from the action itself. We find that the equations of motion generated by both methods agree, but the EFT approach yields certain advantages. To demonstrate some of these advantages we construct a quadratic action that is valid to very small distance scales---much smaller than the scales at which ordinary hydrodynamic theories break down. Such an action captures the full thermodynamic and quantum behavior of reactions and diffusion at quadratic order. Finally, taking the low-frequency and low-wavenumber limit, we reproduce the  linearized version of the well-known reaction-diffusion equations as a final coherence check. 

  \end{onecolabstract}
]
\saythanks


\
{
  \hypersetup{linkcolor=blue}
  \tableofcontents
}

\section{Introduction}\label{Intro}

The long-distance and late-time dynamics of reactive fluid flows are most often studied numerically at the level of the non-relativistic equations of motion. The standard derivations of the equations of motion are rather cumbersome and require extensive uses of thermodynamics~\cite{reaction book}. Further, the usual formulation is incapable of systematically accounting for quantum and thermal fluctuations. In this paper, we present an alternative approach based on effective field theory (EFT), which presents us with certain advantages. In particular, from the EFT perspective, the infrared (IR) degrees of freedom of the system are fully captured by a local action and the field content consists of Goldstones and pseudo-Goldstones~\cite{Nicolis,Zoology,More gapped Goldstones,coset,Finite T superfluids}. As a result, the EFT can be fully specified by symmetry considerations up to finitely many experimentally-determined parameters at any given order in the derivative and field expansions. As a result, the EFT approach requires symmetry as the only input. 

Since reactive fluid flows take place at finite temperature, we must formulate our EFT using the in-in formalism on the Schwinger-Keldysh (SK) contour. As a result, the field content is doubled. In recent years, much progress has been made in understanding non-equilibrium systems---that is, systems out of thermodynamic equilibrium---from the perspective of EFT~\cite{Landry,H. Liu 1,H. Liu 2,H. Liu 3,H. Liu 2.2,H. Liu 2.3,H. Liu 2.1,H. Liu 4,FD 1,Harder,Banerjee,Jensen,Kovtun,Grozdanov,Haehl,Haehl 2,Fluid Manifesto,Hongo 1,D.V. Volkov,Hongo 2,Hongo 3,Hongo 4}. In particular, EFTs may be defined on the SK contour and can systematically account for dissipation as well as quantum and thermal fluctuations. Since these actions can be constructed entirely from symmetry considerations, the (local) laws of thermodynamics follow directly from the equations of motion and do not need to be included as extra inputs. In particular, the second law of thermodynamics, namely that the divergence of the entropy current is nonnegative $\partial_\mu s^\mu\geq0$, can be derived at the level of the classical equations of motion~\cite{H. Liu 2}.


In this paper, we will use both the more standard equations-of-motion approach and the EFT approach to hydrodynamics to develop a relativistically-correct theory of reactions that take place in fluid phase. Our EFT approach is as follows. In keeping with the usual understanding of hydrodynamics, we suppose that all relevant information about fluids is captured by correlations among the stress-energy tensor and particle number currents. Since we will be interested in a fluid with multiple particle species, we include $N$-many $U(1)$ currents---each species of particle corresponds to a different $U(1)$ charge. Then, to allows for reactions, we explicitly break certain linear combinations of these $U(1)$ symmetries. In order that our EFT provide a reliable description of chemical reactions, we must require that the characteristic time-scale of reactions is much longer than the collision time of the molecules in the fluid. From an EFT perspective, this separation of scales is imposed by requiring that the explicitly broken symmetries are realized as approximate symmetries. Finally in fluid phase, the hydrodynamic modes corresponding to unbroken conserved charges possess a kind of gauge symmetry known as a chemical shift~\cite{Nicolis}. We postulate that the hydrodynamical modes corresponding to approximate $U(1)$ charges also possess these chemical shift symmetries. We find that using this EFT approach, we recover the same equations of motion that we obtained using the more conventional method involving conserved currents and constitutive relations. This gives us confidence that our EFT approach---as well as the symmetries we postulate---are correct.

The EFTs constructed in this paper have a wide range of applications including condensed matter physics, nuclear physics, and large-scale cosmology. In addition to having practical applications, such EFTs are of theoretical interest because, unlike ordinary hydrodynamics, there is no non-trivial regime in which entropy does not increase. Thus, even in the leading-order hydrodynamical limit, entropy is always produced. 

The structure of this paper is as follows. In \S\ref{EOM}, we begin by reviewing the derivation of leading-order hydrodynamics with multiple species of independently conserved particles from conservation equations and constitutive relations. Then to allow for the possibility of chemical reactions, we relax certain particle-number conservation equations and find leading order hydrodynamic equations for chemical reactions taking place in fluid phase. In \S\ref{DC EFT}, using the non-equilibrium EFT framework developed in \cite{H. Liu 1,H. Liu 2,H. Liu 3}, we construct an EFT description to describe chemical reactions in fluid phase and reproduce the results of \S\ref{EOM}. In \S\ref{reaction diffusion section}, we demonstrate some advantages of the EFT approach by constructing a quadratic action valid to all orders in the derivative expansion that describes the reactive and diffusive behavior of the system.  Finally in \S\ref{Conclusions}, we discuss the comparative advantages of the EFT program over the more standard $\text{$\text{(non-)}$}$conservation equation approach.

Throughout this paper, we will use the mostly plus convention $\eta_{\m\n}=\text{diag} (-,+,+,+)$. 

\section{The equations of motion approach}\label{EOM}
 
 Our ultimate goal will be to construct an EFT of reactive fluids. But as a first step, it his helpful to build some physical intuition by deriving the equations of motion from the conservation equations for energy and momentum as well as the $\text{(non-)}$conservation equations for particle numbers of the various species. We begin by constructing the standard relativistic hydrodynamical equations of motion for a fluid with $N$ conserved species of particles and then show how it can be modified to account for chemical reactions. We will find that entropy can be produced even at leading order in the derivative expansion and we provide a thermodynamic understanding of this entropy production. 

\subsection{Without chemical reactions} \label{EOM without}
 
As a warm-up let us consider the hydrodynamics of a fluid with multiple conserved species of particles. This is a straightforward generalization of the discussion found in \cite{Weinberg cosmo}. The hydrodynamic equations merely express the conservation of whatever quantities do not change in time. By Noether's theorem, we know that each conserved quantity corresponds to a symmetry of the microscopic system. In relativistic systems, the microscopic theory is Poincaré-invariant; that is, it is invariant under spacetime translations and Lorentz transformations as well as any internal symmetries. We will consider system in which the internal symmetry group is $[U(1)]^N$, corresponding to the independent conservation of $N$ species of particles. 

From Poincaré-invariance, Noether's theorem requires the conservation of the stress-energy tensor 
\begin{equation}\label{energy conservation}\partial_\mu T^{\mu\nu}(x)=0,\end{equation}
where $T^{\mu\nu}$ is symmetric, as well as the conservation of $\cM^{\mu\nu\rho}=x^\mu T^{\nu\rho}-x^{\nu}T^{\mu\rho}$. However, the conservation of $\cM^{\mu\nu\rho}$ follows directly from the symmetry and conservation of $T^{\mu\nu}$ and therefore contains no additional information; we will therefore ignore it. Additionally, for each internal $U(1)$ symmetry, there is a corresponding conserved current 
\begin{equation}\label{particle conservation}\partial_\mu J^{A\mu}(x)=0,\end{equation}
where $A=1,\dots,N$. 

We are interested in systems in local equilibrium. Therefore, we expect that it is possible to express $T^{\mu\nu}$ and $J^{A\mu}$ in terms of thermodynamic quantities that are promoted to local functions of space and time. A system in thermal equilibrium is described by the four-velocity $u^\mu$, which picks out the rest-frame of the system, as well as the temperature $T$ and the chemical potentials $\mu^A$, corresponding to the conserved currents $J^{A\mu}$. Letting $\Delta^{\mu\nu}=\eta^{\mu\nu}+u^{\mu}u^{\nu}$ be the projection operator onto the orthogonal subspace of $u^\mu$, we have
\begin{equation}\begin{split}\label{constitutive relations}T^{\mu\nu}&=\cE u^\mu u^{\nu}+\cP \Delta^{\mu\nu} +(q^\mu u^\nu+q^\nu u^\mu)+t^{\mu\nu }\\
J^{A\mu} &= \cN^A u^\mu + j^{A\mu},
 \end{split}\end{equation}
where $\cE$, $\cP$, and $\cN^A$ are scalars; $q^\mu$  and $j^{A\mu}$ are vectors orthogonal to $u^\mu(x)$; and $t^{\mu\nu}$ is a symmetric, traceless  tensor. Moreover, all of these quantities depend on $T(x)$ and $\mu^A(x)$.

Hydrodynamics is organized as a derivative expansion; leading-order hydrodynamics involves ignoring all terms of the expressions of (\ref{constitutive relations}) that involve derivatives. Thus, the only vector quantity is $u^\mu$ and the only tensor quantities are $u^\mu u^\nu$ and $\Delta^{\mu\nu}$. Hence at leading order, we have
\begin{equation}\begin{split}\label{constitutive relations 0}T^{\mu\nu}&=\epsilon u^\mu u^{\nu}+p \Delta^{\mu\nu}\\
J^{A\mu} &= n^A u^\mu,
\end{split}\end{equation}
where we have identified $\cE$ with the local energy density $\epsilon(x)$, $\cP$ with the local pressure $p(x)$, and $\cN^A$ with the local particle number densities $n^A(x)$. The equilibrium equation of state provides the equation $p=p(T,\mu^A)$, which can be used to compute the energy density~$\epsilon$, entropy~$s$, and particles numbers~$n^A$ by $\epsilon+p=Ts+\mu^A n^A$, where $s=\partial p/\partial T$ and $n^A=\partial p/\partial\mu^A$. Using the longitudinal component of (\ref{energy conservation}) and the particle-conservation equations (\ref{particle conservation}), we have, at leading order in the derivative expansion,
\begin{equation}\begin{split}\label{0 order conservation}\partial_\mu((\epsilon+p)u^\mu)&=u^\mu\partial_\mu p
\\ \partial_\mu(n^A u^\mu)&=0.
\end{split}\end{equation}
By using the thermodynamic relations $\epsilon+p=Ts+\mu^A n^A$ and $dp=sdT+n^A d\mu^A$ in conjunction with (\ref{0 order conservation}), we find that the entropy current is conserved
\begin{equation}\partial_\mu s^\mu=0,\end{equation}
where $s^\mu\equiv s u^\mu$. Thus, leading-order hydrodynamics is non-dissipative. We will find that this is not so if we allow chemical reactions to take place. 

\subsection{With chemical reactions} \label{EOM with}

Now, consider a fluid with $N$ species of particles that  may undergo chemical reactions with one another. In this case, particle number is no longer conserved, so (\ref{particle conservation}) no longer holds; however, there may be certain linear combinations of the non-conserved $U(1)$ currents that are conserved. To see how this is the case, let $\text X^A$, for $A=1,\dots,N$, represent the species of particle corresponding to $J^{A\mu}$ and suppose that the following $k\leq N$  chemical reactions are allowed
\begin{equation}\begin{split}\label{chemical reactions}C_1^A\text X^A & \longleftrightarrow {C'}_1^A \text X^A
 \\ &~~~\vdots 
 \\ C_k^A\text X^A & \longleftrightarrow {C'}_k^A \text X^A,\end{split}\end{equation}
where the $C$'s are positive, real coefficients and repeated indices are summed over. 
The double arrows indicate that the system is near equilibrium, so the reactions may go in either direction. Any linear combination of charges $Q^A\equiv \int d^3 x~J^{A0}(x)$ that is preserved by all of the above chemical reactions is a genuinely conserved charge; any linear combination of $Q^A$ that is not preserved is not conserved. The most general linear combination of charges is $q^A Q^A$, for real coefficients $q^A$, were we have used the convention that repeated indices are summed over. Then assuming without loss of generality that each $\text X^A$ has unit charge with respect of $Q^A$ and zero charge with respect to $Q^B$ for $B\neq A$, the composite charge $ q^A Q^A$ is conserved if and only if 
\begin{equation}\begin{split}C_1^A q^A & = {C'}_1^A q^A
 \\ &~~~\vdots 
 \\ C_k^Aq^A & = {C'}_k^A q^A,\end{split}\end{equation}
which can be expressed in the matrix equation $c^A_i q^A=0$, for $c^A_i\equiv C^A_i- C'^A_i$. 

Let $\cP^A_B$ be the projection operator onto the null-space of $c^A_i$. Then $\cP^A_B$ projects onto the subspace of conserved charges. The $\text{(non-)}$conservation equations for the particle-number currents are therefore
\begin{equation}\label{particle non-conservation} \partial_\mu J^{A\mu}(x)=\Gamma^A(x),\end{equation}
where $\cP^A_B\Gamma^B=0$ and $\Gamma^A$ is a local functional of $T(x)$ and $\mu^A(x)$. It follows immediately that $\cP^A_B J^{B\mu}$ are conserved. The presence of chemical reactions does not affect Poincaré symmetry, so the conservation of the stress-energy tensor, (\ref{energy conservation}) still holds. 

Following the ordinary hydrodynamics example, we parameterize the stress-energy tensor and currents according to (\ref{constitutive relations}). And at leading order in the derivative expansion, we have (\ref{constitutive relations 0}). As before, we identify $\cE$, $\cP$, and $\cN^A$  with the local energy density $\epsilon$, pressure $p$, and particle numbers $n^A$, respectively. Then, using the longitudinal component of (\ref{energy conservation}) as well as (\ref{particle non-conservation}), we have, at leading order in the derivative expansion,
\begin{equation}\begin{split}\label{0 order non-conservation}\partial_\mu((\epsilon+p)u^\mu)&=u^\mu\partial_\mu p
\\ \partial_\mu(n^A u^\mu)&=\Gamma^A.
\end{split}\end{equation}
By using the thermodynamic relations $\epsilon+p=Ts+\mu^A n^A$ and $dp=sdT+n^A d\mu^A$ in conjunction with (\ref{0 order non-conservation}), we find that the entropy current is not conserved
\begin{equation}\label{2nd law eom}\partial_\mu s^\mu=-\frac{\mu^A}{T}\Gamma^A.\end{equation}
To ensure that the second law of thermodynamics is satisfied, we require that the r.h.s. is always nonnegative. Notice that unlike ordinary hydrodynamics, if we allow chemical reactions, then entropy can be produced even at leading order.

\subsection{The meaning of $\partial_\m s^\m$}\label{EOM thermo}

Now that we have an expression for the entropy production in a reacting perfect fluid, we would like to understand its physical meaning. That is, from a thermodynamic standpoint, what is the meaning of the terms on the r.h.s. of (\ref{2nd law eom})? Imagine that we have a fluid element with particle numbers $N^A$, energy $E$, and volume $V$. 
 To describe a perfect fluid, we assume that the fluid elements do not exchange any net particles with one another and energy is exchanged through work alone. Further, in order to have a sensible derivative expansion, we must require a separation of time scales: all chemical reactions must take place on time scales much longer than the collision time. Therefore the temperatures and pressures of each species of particle within a given volume element must be equal. As a result, no entropy can be produced through either energy or volume exchanges among particle species within a given volume element. Thus, the change in entropy with respect to time in the local rest-frame of a given volume-element is 
 \begin{equation}\begin{split} \dot S &= \frac{\partial S}{\partial E} \dot E+\frac{\partial S}{\partial V} \dot V+\frac{\partial S}{\partial N^A} \dot N^A
 \\ &=\frac{1}{T} \dot E-\frac{p}{T} \dot V -\frac{\mu^A}{T} \dot N, \end{split}\end{equation} 
where the second equality makes use of standard thermodynamic identities. 
Since no net particles can be exchanged among the volume-elements, the only way for $N^A$ to change in time is through chemical reactions, meaning that $\dot N^A/V=\Gamma^A$. And since energy is exchanged among volume elements via work alone, we have $\dot E = p \dot V$. Then, using covariant notation such that $\dot S/V\to\partial_\m s^\m$, we have
 \begin{equation} \partial_\m s^\m = -\frac{\m^A}{T} \Gamma^A,  \end{equation} 
 which matches the result (\ref{2nd law eom}) of the previous subsection.

\subsection{An explicit example}

All of the discussion surrounding chemical reactions and $\text{(non-)}$conserved currents so far has been rather abstract and formal. In this section, we will build intuition by considering a simple, concrete example. Suppose we have a system that consists of hydrogen, oxygen, and water and that we are working in a temperature and pressure regime such that chemical reactions among these substances are happening in both directions
\begin{equation}2H_2+O_2 \longleftrightarrow 2 H_2 O.\end{equation}
Let $Q_{H_2}$ and $Q_{O_2}$ be the number operators of hydrogen and oxygen molecules respectively, and let $Q_{H_2O}$ be the number operator for water molecules. Notice that because these substances can react, none of these number operators correspond to a conserved quantity. However, since we have three species of particles and one allowed chemical reaction, we expect that there should be exactly two conserved charges.  In particular, notice that whenever a chemical reaction occurs, the {\it total} number of hydrogen and oxygen atoms never changes. Let $N_H$ and $N_O$ be the number operators that count the {\it total} number of hydrogen and oxygen atoms, respectively. Then, we have that
\begin{equation}N_H = 2 Q_{H_2}+2 Q_{H_2O},~~~~~ N_O = 2 Q_{O_2} + Q_{H_2O}\end{equation}
are exactly conserved charges. 

Let us work in the $N_H$, $N_O$, and $Q_{H_2 O}$ charge basis and let $J^\mu_{H}$, $J^\mu_{O}$, and $J^\mu_{H_2O}$ be the corresponding currents. We then have that 
\begin{equation}\partial_\mu J^\mu_{H}=\partial_\mu J^\mu_{O}=0,\end{equation}
and we define $\Gamma_{H_2O}$ such that
\begin{equation}\label{water current}\partial_\mu J^\mu_{H_2O}=\Gamma_{H_2O}.  \end{equation}
The divergence of the entropy current is then 
\begin{equation} \partial_\mu s^\mu = -\frac{\mu_{H_2O}}{T} \Gamma_{H_2O}. \end{equation}
To build intuition about how the second law of thermodynamics is satisfied, consider three cases:
\begin{itemize}
\item \underline{$\mu_{H_2 O}>0$:} Then, it is energy-favorable for the number of of water molecules to decrease, meaning that $\partial_\mu J^\mu_{H_2O} = \Gamma_{H_2O}<0$. As a result, $\partial_\mu s^\mu >0$. 
\item  \underline{$\mu_{H_2 O}<0$:} Then, it is energy-favorable for the number of water molecules to increase, meaning that $\partial_\mu J^\mu_{H_2O} = \Gamma_{H_2O}>0$. As a result, $\partial_\mu s^\mu >0$. 
\item \underline{$\mu_{H_2 O}=0$:} Then, $\partial_\mu s^\mu =0$. Notice that because $Q_{H_2O}$ is not conserved, in equilibrium, $\mu_{H_2O}=0$. Thus no entropy is produced in equilibrium, as expected. 
\end{itemize}
In all of these cases, the second law of thermodynamics is manifestly satisfied.

\section{The non-equilibrium EFT approach}\label{DC EFT}

We begin by reviewing some of the basics of non-equilibrium EFT; for a very nice and in-depth review, consult \cite{H. Liu 1}.  These EFTs describe systems out of finite-temperature equilibrium. To account for the fact that their equilibrium density matrix is of the form
\begin{equation}\rho = \frac{e^{-\beta_0 (H-\m Q)}}{\text{tr}\squ{ e^{-\beta_0 (H-\m Q)}}},\end{equation}
which can never be a pure state for finite inverse temperature $\beta_0$, we must perform all computations using the in-in formalism on the Schwinger-Keldysh (SK) contour. Ordinarily, a pure state requires one copy of the time-evolution operator to evolve it in time, but density matrices, being linear operators, require two copies of the time-evolution operator. Thus, the sources in this formalism are doubled---one copy for each time evolution operator---and the generating functional takes the form
\begin{equation}e^{W[\cJ_1,\cJ_2]} = \text{tr}\squ{U(+\infty,-\infty;\cJ_1) \rho U^\dagger(+\infty,-\infty;\cJ_2)},\end{equation}
where $U(+\infty,-\infty;\cJ_s)$ is the time-evolution operator from the distant past to the distant future in the presence of source $\cJ_s$, for $s=1,2$. 

We would like to find some sort of effective action that can give rise to the generating functional $W[\cJ_1,\cJ_2]$, that is\footnote{The subscript SK indicates that we impose SK boundary conditions, namely that in the distant future, the two copies of the fields are equal $\varphi_1(+\infty)=\varphi_2(+\infty)$.}
\begin{equation}e^{W[\cJ_1,\cJ_2]} = \int_\text{SK} \cD [\varphi_1  \varphi_2] ~e^{i I_\text{EFT}[\varphi_1,\varphi_2;\cJ_1,\cJ_2]}, \end{equation}
for some infrared (IR) fields $\varphi_s$ for $s=1,2$. We call $I_\text{EFT}$ the non-equilibrium effective action. Notice that because the generating functional has doubled sources, the effective action has doubled field content as well.   Following the usual EFT philosophy, it is our goal to express $I_\text{EFT}$ as a linear combination of all terms compatible with symmetries with finitely many unknown coefficients at any given order in the derivative and field expansions. But first, there are several rules that all non-equilibrium actions must satisfy~\cite{H. Liu 1}. We present them below without proof. 
\begin{itemize}
\item The non-equilibrium effective action contains terms that involve products of 1 and 2 fields; that is, the fields on each leg of the SK contour can interact with one another.
\item Terms of $I_\text{EFT}$ are complex. Unitarity imposes the following constraints 
\begin{equation}\begin{split}\label{unitarity constraints} I^*_\text{EFT} [\varphi_1,\varphi_2;\cJ_1,\cJ_2] & = - I_\text{EFT} [\varphi_2,\varphi_1;\cJ_2,\cJ_1]
\\ \text{Im} I_\text{EFT} [\varphi_1,\varphi_2;\cJ_1,\cJ_2] &  \geq 0, \text{for any }\varphi_{1,2}, \cJ_{1,2}
\\ I_\text{EFT} [\varphi_1=\varphi_2;\cJ_1=\cJ_2] & = 0. 
\end{split}\end{equation} 
\item Any symmetries of the UV action are symmetries of the effective action except for symmetries that involve time-reversal.
\item For systems at finite temperature, time-reversing symmetries manifest as the so-called dynamical KMS symmetries, which can be derived from the KMS conditions of thermal partition functions \cite{H. Liu 1}. Suppose that $\Theta$ represents some time-reversing symmetry of the UV theory. The non-equilibrium effective action is not invariant under the same time-reversing symmetries that exist in the UV; this allows the production of entropy. Instead, the non-equilibrium effective action is invariant under
\begin{equation}\begin{split} \varphi_1(x)&\to\Theta\varphi_1(t-i\theta,\vec x),
\\ \varphi_2(x)&\to \Theta \varphi_2(t+i(\beta_0-\theta),\vec x),
 \end{split}\end{equation} 
 for any $\theta\in[0,\beta_0]$, where $\beta_0$ is the inverse equilibrium temperature. It may seem odd that such symmetries require non-local transformations, but in the classical limit, they become local. In order to take the classical limit it is helpful to define the symmetric and anti-symmetric fields by
 \begin{equation}\label{ra basis}\varphi_r \equiv \frac{1}{2}\cur{\varphi_1+\varphi_2},~~~~~~~~~~\varphi_a \equiv \varphi_1-\varphi_2 .\end{equation}
Then the classical dynamical KMS symmetries act by
\begin{equation}\label{classical dynamical KMS}\varphi_r(x)\to\Theta\varphi_r(x),~~~~~\varphi_a(x)\to\varphi_a(x)+\Theta [i\beta\partial_0\varphi_r(x)]. \end{equation}
It turns out that the $r$-type variables behave like classical fields and $a$-type variables describe the thermal and quantum fluctuations; see Appendix \S\ref{a-type variables appendix}. Notice that the change in $\varphi_a$ under the classical dynamical KMS symmetry depends on the derivative of $\varphi_r$. Thus, it is natural to consider $\varphi_a$ and $\partial_0 \varphi_r$ as contributing at the same order in the derivative expansion. 
\end{itemize}

Finally, applying a Noether-like procedure to the classical dynamical KMS symmetries, it is possible to construct a current $s^\mu$ whose divergence on-shell is always nonnegative. This current can be identified with the local entropy current, and the nonnegative gradient enforces the local statement of the second law of thermodynamics\cite{H. Liu 2}. Consider the effective action without sources in the classical limit $I_\text{EFT}[\varphi_r,\varphi_a]=\int d^4 x ~\cL _\text{EFT}[\varphi_r,\varphi_a]$. To keep things fully general, suppose the classical dynamical KMS transformations are
\begin{equation}\begin{split}\varphi_r(x)&\to\Theta\varphi_r(x)
\\ \varphi_a(x)&\to\Theta\varphi_a(x)+i\Theta\Lambda_r(x),
\end{split}\end{equation}
for some $r$-type field $\Lambda_r$. 
Then under a dynamical KMS transformation, the effective Lagrangian can change by at most a total derivative, $\cL_\text{EFT}\to\cL_\text{EFT}+\partial_\mu V^\mu$. We can expand $V^\mu$ in powers of $a$-type fields
\begin{equation}V^\mu=iV^\mu_0+V^\mu_1+\cdots,\end{equation}
where $V^\mu_k$ contains $k$ factors of $a$-type fields. Since the dynamical KMS symmetry is discrete and terms of $\cL_\text{EFT}$ must all have at least one $a$-type field, it is possible (using integration by parts) to define $\cL_\text{EFT}$ such that $V^\mu=iV^\mu_0$. However, if we express $\cL_\text{EFT}$ in such a way that the terms linear in $a$-type fields have no derivative acting on the $a$-type fields, then we may have a non-zero $V_1^\mu$ but still have $V^\mu_k=0$ for $k\geq 2$. In this case, define the entropy current by
\begin{equation}s^\mu=V_0^\mu -\hat V_1^\mu,\end{equation}
where $\hat V^\mu_1=V^\mu_1|_{\varphi_a=\Lambda_r}$. Using the fact that $\text{Im} I_\text{EFT} \geq 0$ it has been demonstrated in \cite{H. Liu 2} that $\partial_\mu s^\mu\geq 0$, meaning that the second law of thermodynamics is automatic.

\subsection{Without chemical reactions} 

Let us construct the effective action for a system that, in addition to Poincaré symmetry, has $N$-many conserved charges. This corresponds to a fluid with $N$ species of particles that cannot undergo chemical reactions. We start from the assumption that the long-distance and late-time dynamics of such a system are described entirely by correlations among the conserved quantities. Thus, the generating functional we are interested in is 
\begin{equation}\begin{split} e^{W[g_{1\m\n}, g_{2\m\n}, \cA^A_{1\m},\cA^A_{2\m} ]} =   \text{tr}\big[U(+\infty,-\infty;g_{1\m\n},\cA^A_{1\m})
\\ \times \rho U^\dagger(+\infty,-\infty; g_{2\m\n},\cA^A_{2\m})\big] ,\end{split} \end{equation} 
where functional derivatives with respect to the background metrics $g_{s\m\n}$ give correlators among the stress-energy tensors and functional derivatives with respect to the $U(1)$ background gauge fields $\cA^A_{s\m}$ for $A=1,\dots,N$ give correlators among the $U(1)$ currents. Notice that because the stress-energy tensor and the $U(1)$ currents are all conserved and the sources are doubled, the generating functional must be invariant under two copies of diffeomorphism symmetries as well as two copies of the $N$-many $U(1)$ gauge symmetries. 
 Using the Stückleberg tricks of \cite{H. Liu 1}, we promote these gauge transformations to dynamical fields and thereby `integrate in' the hydrodynamical modes of the non-equilibrium EFT. We have
 \begin{equation}\begin{split} e^{W[g_{1\m\n}, g_{2\m\n}, \cA^A_{1\m},\cA^A_{2\m} ]} & \\= \int \cD [X^\m_s \pi_s^A ]&~e^{I_\text{EFT} [G_{1MN}, G_{2MN}, \cB^A_{1M},\cB^A_{2M } ]}, \end{split} \end{equation} 
 where now the effective action is defined on fluid worldvolume coordinates $\phi^M$ for $M=0,1,2,3$  and  
 \begin{equation}\begin{split}\label{fluid metrics and gauge fields}G_{sMN} (\phi) &\equiv \frac{\partial X_s^\m(\phi)}{\partial\phi^M} g_{s\m\n} (X_s(\phi))\frac{\partial X_s^\n(\phi) }{\partial\phi^N},
 \\ \cB^A_{sM}(\phi)&\equiv \cA^A_{s\m}(X_s(\phi))\frac{\partial X_s^\m(\phi)} {\partial\phi^M}+\frac{\partial \pi^A_s(\phi) }{\partial\phi^M},
\end{split} \end{equation} 
where $X^\m_s(\phi)$ and $\pi^A_s(\phi)$ for $s=1,2$ are the dynamical fields.
We interpret $X^\m_s(\phi)$ as embeddings of the fluid worldvolume into the physical spacetime. Then $G_{sMN}$ can be thought of as the pull-back metrics onto the fluid worldvolume and $\cB_{sM}^A$ can be thought of as the $U(1)$ gauge fields expressed in `unitary gauge.' 

In fluid phase, there are no spontaneously broken symmetries, so following the philosophy of \cite{Landry}, we have that $\pi^A_s$ possess gauge symmetries, known as `chemical shifts'~\cite{Nicolis}
\begin{equation}\label{chemical shift symmetry} \pi^A_s\to\pi^A_s+ f^A(\phi^I) ,\end{equation} 
for arbitrary spatial functions $f^A(\phi^I)$ for $I=1,2,3$. 
Additionally, the effective action is invariant under a reduced diffeomorphism invariance on the coordinates $\phi^M$, which we will refer to as the `fluid symmetries' and are give by 
\begin{equation}\label{fluid diff symmetry} \phi^M\to\phi^M +\xi^M(\phi^I), \end{equation} 
for arbitrary spatial functions $\xi^M(\phi^I)$, where $M=0,1,2,3$ and $I=1,2,3$. Thus, the building-blocks for the effective action at leading order in the derivative expansion are as follows. We find it convenient to use the $r,a$-basis (\ref{ra basis}). The $r$-type building-blocks are respectively the local inverse-temperature four-vector and local chemical potentials
\begin{equation}\label{r-type building blocks}\beta^\m\equiv \beta_0 \frac{\partial X^\m_r}{\partial\phi^0},~~~~~~~~~~\mu^A\equiv u^\m\partial_\m\psi_r^A, \end{equation}
where $u^\m\equiv \frac{1}{\beta}\beta^\m$ is the local fluid four-velocity, $\beta\equiv \sqrt{-\beta^\m\beta_\m}$ is the local inverse temperature, and $\psi_r^A\equiv \mu_0^AX_r^t+\pi^A_r$ such that $\mu_0^A$ are the equilibrium chemical potentials. Additionally, the $a$-type building blocks are\footnote{We define $\partial_\mu\equiv \partial/\partial X_r^\mu$. }
\begin{equation}\label{a-type building blocks}\partial_\mu\pi_a^A,~~~~~~~~~~\partial_\mu X_a^\nu.\end{equation}
Performing a change of coordinates to the physical spacetime $x^\mu\equiv X_r^\mu$, the classical Lagrangian at lowest order in derivatives is 
\begin{equation}\cL_\text{EFT} =T^{\mu\nu}\partial_\mu X_{a\nu}+J^{A\mu}\partial_\mu\pi_a^A,\end{equation}
where repeated indices are summed over and 
\begin{equation}T^{\mu\nu}=\epsilon(\beta,\mu^A)u^\mu u^\nu+p(\beta,\mu^A) \Delta^{\mu\nu}\end{equation}
is the stress-energy tensor and
\begin{equation}J^{A\mu}=n^A(\beta,\mu^A) u^\mu\end{equation}
are the Noether currents associated with conserved $U(1)$ charges $Q_A$. The equations of motion are found by varying the $a$-type fields and then setting $a$-type fields to zero. The equations of motion are therefore the conservation equations 
\begin{equation}\begin{split} \partial_\m T^{\mu\nu}  = 0,~~~~~ \partial_\mu J^{A\mu}  = 0.
 \end{split}\end{equation}
Lastly, we must impose the (classical) dynamical KMS conditions since the equilibrium state of a fluid is necessarily thermal equilibrium. Suppose that $\Theta$ represents some symmetry of the microscopic theory that involves time inversion. At the classical level, the action of the dynamical KMS transformation on $r$-variables is equivalent to the action of $\Theta$. The action on $a$-type variables, however, is more interesting~\cite{H. Liu 1,H. Liu 2,H. Liu 3}:
\begin{equation}\begin{split}\label{D-KMS} X_a^\mu(\phi) &\to \Theta{ X_a^\mu(\phi)} -i\Theta {\beta^\mu(\phi)}+i\beta_0\delta^\m_0,
\\ \pi_a^A(\phi)&\to\Theta\pi_a^A(\phi)+i\Theta\squ{\beta_0\partial_0\pi_r^A(\phi)}.
\end{split}\end{equation}
Imposing that $\cL_\text{EFT}$ transforms by a total derivative under (\ref{D-KMS}) yields the following relations
\begin{equation}\begin{split} \label{therm rel}\epsilon+p&=-\beta\frac{\partial p}{\partial\beta}+\mu^A\frac{\partial p}{\partial\mu^A},
\\ n^A&=\frac{\partial p}{\partial\mu^A},
\end{split}\end{equation}
which are the standard thermodynamic relations. Thus the leading-order stress-energy tensor and particle number currents take the expected form. And since the equations of motion are just the conservation of the stress-energy tensor and particle number currents, we have reproduce the leading-order hydrodynamic equations with multiple conserved charges. 

The dynamical KMS transformation is a discrete symmetry that classically only acts non-trivially on $a$-type variables. But because our effective action is linear in $a$-type variables, it inherits an accidental, continuous $U(1)$ symmetry. As a result, Noether's theorem furnishes a conserved current for leading-order hydrodynamics given by
\begin{equation}s^\mu=p\beta^\mu-\beta_\nu T^{\mu\nu}-\beta\mu^A J^{A\mu}.\end{equation}
We identify this current with the entropy current, which is conserved in the perfect-fluid limit.

\subsection{With chemical reactions}

Now we construct the effective action for a system that can undergo chemical reactions. Such a system is invariant under Poincaré symmetry, but now only has $k<N$-many conserved $U(1)$ charges and $(N-k)$-many approximately conserved $U(1)$ charges; see \S\ref{EOM with}. 
Let $\cP_B^A$ be the projection operator onto the space of conserved charges. 
Just as in the previous subsection, we wish to include source-terms for all of the hydrodynamical modes, so we include the metrics $g_{s\m\n}$ as source-terms for the stress-energy tensor. Unfortunately, now some of the hydrodynamical modes do not correspond to exactly conserved quantities, so introducing gauge fields for these approximate symmetries is not possible. We circumvent this problem as follows. Let $S[\Psi]$ be the UV action. Introduce an auxiliary field $\kappa(x)$ and define the new action $S'[\Psi,\kappa]$. We allow $\kappa$ to transform in such a way that this new action is invariant under $N$-many $U(1)$ symmetries and such that
\begin{equation}S'[\Psi,\kappa=\kappa_0]=S[\Psi],\end{equation}
for some fixed constant $\kappa_0$. Now that $S'$ possesses exact $[U(1)]^N$-symmetry, we may introduce gauge fields $\cA_{s\m}^A$ for $A=0,\dots,N$. Thus our generating functional depends on three kinds of source terms: the metrics $g_{s\m\n}$, the $U(1)$ gauge fields $\cA_{s\m}^A$, and the auxiliary fields $\kappa_s$. It takes the form
\begin{equation}\begin{split}  e^{W[g_{1\m\n}, g_{2\m\n}, \cA^A_{1\m},\cA^A_{2\m},\kappa_1,\kappa_2 ]} =   \text{tr}\big[ U(+\infty,-\infty;g_{1\m\n},\cA^A_{1\m},\kappa_1)
\\ \times  \rho U^\dagger(+\infty,-\infty; g_{2\m\n}.\cA^A_{2\m},\kappa_2) \big].\end{split}\end{equation} 
Then, just as in the previous subsection, we can `integrate in' the hydrodynamic modes of the non-equilibrium EFT. We have
 \begin{equation}\begin{split} e^{W[g_{1\m\n}, g_{2\m\n}, \cA^A_{1\m},\cA^A_{2\m},\kappa_1,\kappa_2 ]} = \int \cD [X^\m_s  \pi_s^A] ~~~~~~~~~~~~~
 \\ ~\times e^{I_\text{EFT} [G_{1MN}, G_{2MN}, \cB^A_{1M},\cB^A_{2M },\cK_1,\cK_2 ]}, \end{split}\end{equation} 
where $G_{sMN}$ and $\cB^A_{sM}$ are given by (\ref{fluid metrics and gauge fields}) and the $\cK_s$ are the `Stückelberged' versions of $\kappa_s$; explicitly, 
\begin{equation} \cK_s(\phi) = e^{i\hat\pi_s^A(\phi) Q_B}\cdot\kappa_s(\phi), \end{equation}
where $\hat \pi_s^A$ are the fields associated with the approximate $U(1)$ symmetries and therefore satisfy $\cP^A_B \hat \pi_s^B=0$. 
Just as in the previous subsection, we require that the effective action must be invariant under the chemical shift gauge symmetry (\ref{chemical shift symmetry}) and the fluid diffeomorphism symmetry (\ref{fluid diff symmetry}). In particular, we postulate that the pseudo-Goldstones corresponding to approximate $U(1)$ symmetries still enjoy chemical shift symmetries. Thus the only difference between the EFTs  that describe chemical reactions and those that do not is the inclusion of $\cK_s$ when chemical reactions are allowed. As a result, the building blocks for the EFT describing chemical reactions are, in addition to (\ref{r-type building blocks}) and (\ref{a-type building blocks}), the $a$-type building-blocks that come from $\cK_s$, namely\footnote{There are no r-type building blocks that arise from $\cK_s$ because the chemical shift gauge symmetries prevent such terms.}
\begin{equation}\hat \pi_a^A\equiv \hat \pi_1^A-\hat \pi_2^A . \end{equation}
Notice that these new $a$-type building-blocks have no derivatives and therefore count at the same order in the derivative expansion as $\partial\pi_r^A$, meaning that they count as first order in the derivative expansion.\footnote{Note that $\pi_a$ is the same order as $\partial\pi_r$ and {\it not} $\partial\psi_r$.} As a result, even at leading order in the derivative expansion, the effective action may depend on terms involving multiple factors of $\hat \pi^A_a$. Thus, the effective Lagrangian at leading order in derivatives is
\begin{equation}\begin{split}\label{chem react EFT} \cL_\text{EFT}=T^{\mu\nu}\partial_\mu X_{a\nu}+J^{A\mu}\partial_\mu\pi_a^A+\cM, \end{split} \end{equation}
where $T^{\m\n}$ and $J^{A\m}$  are the stress-energy tensor and particle-number currents given by
\begin{equation}\begin{split}\label{constitutive relations eft}T^{\mu\nu}&=\epsilon u^\mu u^{\nu}+p \Delta^{\mu\nu} \\
J^{A\mu} &= n^A u^\mu ,
 \end{split}\end{equation}
where $\epsilon,$ $p,$ and $n^A$ are generic functions of  $\beta$, $\mu^A$ and $\cM$ depends on $\beta$, $\mu^A$, and $\hat\pi_a^A$. Recall that $\beta$ and $\mu^A$ are defined in terms of the fields by (\ref{r-type building blocks}). 
To ensure that $\cM$ does not contribute to higher order in the derivative expansion than the other two terms, it can contain terms that are at most quadratic in $\hat \pi_a$. We therefore have
\begin{equation}\cM = \Gamma^A\hat \pi_a^A-\frac{i}{2}\cM_2^{AB} \hat \pi_a^A \hat \pi_a^B,\end{equation}
where $\Gamma^A$ and $\cM_2^{AB}$ may freely depend on $\beta$ and $\mu^A$.
We will see that $\Gamma^A$ is equal to the divergence of the particle number current. 

Notice that the forms of the stress-energy tensor and particle-number currents in the leading-order effective action~(\ref{chem react EFT}) are of the same form as the leading-order constitutive relations~(\ref{constitutive relations 0}). This is a non-trivial check that our EFT is correct; in particular, the fact that we recover the expected form of the stress-energy tensor and particle currents demonstrates that the postulated symmetries (\ref{chemical shift symmetry}-\ref{fluid diff symmetry}) are correct. Suppose, for example, that we did not impose the chemical-shift symmetries for the approximately conserved $U(1)$ fields. Then, we would have covariant building-blocks $\partial_\mu \hat\psi_r^A$, which would indicate that the non-conserved particles could flow independently of the fluid, like in the case of finite-temperature superfluids~\cite{Finite T superfluids}.

To account for the fact that this EFT describes a system at finite temperature, it is necessary to impose the dynamical KMS condition. Imposing this condition will lead to various relations among the terms of $\cL_\text{EFT}$. In particular, since $I_\text{EFT}$ can never have any terms without $a$-type variables, if the action of the dynamical KMS symmetry on $\cL_\text{EFT}$ produces terms without $a$-type variables, they must be total derivatives. We therefore have that
\begin{equation}\label{total derivative 0} T^{\mu\nu}\partial_\mu\beta_\nu+J^{A\mu}\partial_\mu(\beta\mu^A)-i\cM_\star,\end{equation}
must be a total derivative, where $\cM_\star \equiv \cM(\beta,\mu^A,\hat\pi^A_a=i \beta^\m \partial_\m\hat \pi^A_r)$. 
It can be checked that (\ref{total derivative 0}) is a total derivative if and only if
\begin{equation}\begin{split} \epsilon +p&= -\beta \frac{\partial p}{\partial \beta}+\mu^A\frac{\partial p}{\partial\mu^A},
\\ n^A&=\frac{\partial p}{\partial \mu^A},
\\  \cM_\star&=0.
\end{split}\end{equation}
Notice that the first two equations are just the usual thermodynamic relations if we interpret $\epsilon$, $p$, and $n^A$ as the energy density, pressure, and particle number, respectively. The last equation gives the relation 
\begin{equation} \Gamma^A = -\frac{1}{2} \cM_2^{AB} \beta^\m \partial_\m \hat\pi^A_r. \end{equation} 

Now we compute the equations of motion of the system. Equations of motion are found by varying with respect to an $a$-type variable and then setting all remaining $a$-type variables to zero~\cite{H. Liu 1}. Thus, the equations of motion from varying $X_{a\nu}$ are
\begin{equation}\label{energy conservation 00}\partial_\mu T^{\mu\nu}=0,\end{equation}
which are the conservation equations for the stress-energy tensor. The equations of motion from varying $\pi_a^A$ are
\begin{equation}\label{particle non-conservation 00}\partial_\mu J^{A\mu}=\Gamma^A.\end{equation}
Recall that $\cP^A_B \hat\pi_a^B=0$, meaning that  $\Gamma^B \cP_B^A=0$. As a result, we can identify the r.h.s. of this equation with the r.h.s. of (\ref{particle non-conservation}). We have therefore reproduced the leading-order equations of motion for reactive flows that we derived through entirely different means in \S\ref{EOM}. 

Finally, using the dynamical KMS symmetry, we can construct the entropy current. 
In order to calculate the entropy current, we must compute the change in $\cL_\text{EFT}$ under a dynamical KMS transformation. Letting $\Theta \tilde\cL_\text{EFT}$ denote the KMS transformation of the Lagrangian, we have
\begin{equation}\begin{split} \tilde\cL_\text{EFT}\supset i \partial_\mu(p\beta^\mu)+\partial_\mu \big(T^{\mu\nu} X_{a \nu}  +J^{A\mu} \pi_a^A\big).\end{split} \end{equation}
Thus we have that 
\begin{equation}V_0^\mu=p\beta^\mu,~~~~~\hat V_1^\mu=T^{\mu\nu}\beta_{\nu}+J^{A\mu} \beta\mu^A.\end{equation}
Using the identifications that we have constructed with the variables defined in \S\ref{EOM}, we have that $s^\mu =V_0^\mu-\hat V_1^\mu$. 
And using the equations of motion (\ref{energy conservation 00}) and (\ref{particle non-conservation 00}), the divergence of the entropy current is
\begin{equation}\label{2nd law}\partial_\mu s^\mu= -\frac{\mu^A}{T}\Gamma^A,\end{equation}
where $T\equiv 1/\beta$, matching the result of \S\ref{EOM with} and \S\ref{EOM thermo}.

\section{The stochastic reaction-diffusion equation}\label{reaction diffusion section}

We will now use the non-equilibrium EFT machinery developed in the previous sections to derive a generalization of a well-known result. To keep things as simple as possible, suppose that we only want to model the properties of chemical reactions and particle diffusion. In this regime the hydrodynamical fields associated with energy-momentum conservation decouple from those corresponding to particle-number $\text{(non-)}$conservation. Then we may fix the fields $X^\m_s(\phi)=\phi^\m$. Since we already computed the effective action to leading order in derivatives, in this section we will now parameterize the quadratic action to {\it all} orders in the derivative expansion. Such an action will therefore take into account the full statistical and quantum hydrodynamic effects of reactions and diffusion at the linearized level. Since the fluid and physical spacetime coordinates now are equivalent, we label our spacetime by $x^\m$. The residual fluid symmetries acting on these coordinate are then
\begin{equation} x^\m\to x^\m +c^\m,~~~~~~~~~~ x^i\to {R^i}_j x^j, \end{equation} 
for constant $c^\m$ and $R\in\text{SO(3)}$. 
Letting $\psi^A = \mu_0^A t +\pi^A$ for constant $\m_0^A$, the chemical shift symmetries are 
\begin{equation} \pi_r^A\to \pi^A_r +\alpha^A(x^i) ,\end{equation} 
for arbitrary $\alpha$, and the unbroken $U(1)$ symmetries act as 
\begin{equation}\label{shift symmetry r-d}\pi^A _s\to \pi^A_s+ \cP^A_B\lambda^B,\end{equation}
for $s=1,2$ and constant $\lambda^B$. The quadratic action is therefore 
\begin{equation}\cL^{(2)} = \pi_a^A F_{AB} \dot \pi^B_r-\frac{i}{2} \pi_a^A M_{AB} \pi_a^B,  \end{equation} 
where $F_{AB}$ are real functions of $\partial_t$ and $\partial_i\partial^i$. To ensure that $\pi_{r/a}^B\cP_B^A$ correspond to conserved charges, $F_{AB}\cP^A_C$ must vanish for $\omega, \vec k\to0$. Since such an action is non-local in position space and quadratic in the fields, it is helpful to take the Fourier transform 
\begin{equation}\label{Fourier Lagrangian}\cL^{(2)} =-i\omega \bar \pi_a^A F_{AB}(i\omega,k^2)  \pi^B_r-\frac{i}{2} \bar \pi_a^A M_{AB}(i \omega,k^2) \pi_a^B .  \end{equation} 
The full dynamical KMS symmetry requires that $\cL^{(2)}$ be invariant under the following transformations
\begin{equation}\begin{split} \pi_r^A(\omega,\vec k)  \to &e^{\cur{\theta -\frac{\beta_0}{2}}\omega} \bigg(\pi_r^A(-\omega,\vec k) \cosh{\frac{\omega\beta_0}{2}}
\\&~~~~~~~~~~~~~~~~~~+\frac{1}{2} \pi^A_a(-\omega, \vec k) \sinh{\frac{\omega\beta_0}{2}} \bigg)
\\ \pi_a^A(\omega,\vec k)  \to& e^{\cur{\theta -\frac{\beta_0}{2}}\omega} \bigg( \pi_a^A(-\omega,\vec k) \cosh{\frac{\omega\beta_0}{2}}  
\\ & ~~~~~~~~~~~~~~~~~~~+2 \pi^A_r(-\omega, \vec k) \sinh{\frac{\omega\beta_0}{2}} \bigg)
\\ \omega \to& -\omega ,
\end{split} \end{equation}
where $\beta_0$ is the equilibrium temperature and $\theta\in [0,\beta_0)$ is an arbitrary constant. 
This dynamical KMS symmetry gives the constraint that 
\begin{equation}\begin{split} \label{fluctuation dissipation}\frac{\omega}{2}\big(F^{AB}(i \omega,k^2)+& F^{BA}(-i\omega,k^2) \big)\\ &= M^{AB}(i\omega,k^2) \tanh{\frac{\omega \beta_0}{2}},  \end{split} \end{equation} 
or written in matrix notation, $\frac{\omega}{2}(F+F^\dagger) = M \tanh{\frac{\omega\beta_0}{2}} $. 

Now we find the equations of motion. The ordinary equations of motion governing the expectation value of the fields are $-i \omega F^{AB}\pi^B =0$. However, since we have terms quadratic in $a$-type fields, we can do better. In particular we can include noise resulting from thermal and quantum processes \cite{H. Liu 1,H. Liu 2,H. Liu 3}. We have
\begin{equation}\label{Fourier eom} -i\omega F^{AB} \pi^B = \zeta^A , \end{equation}
where $\zeta^A$ is a Gaussian-random stochastic variable with two-point function $\vev{\zeta^A\zeta^B}= M^{AB}$; see Appendix \ref{a-type variables appendix}.  We now see that (\ref{fluctuation dissipation}) is a straight-forward generalization of the usual fluctuation-dissipation theorem given in \cite{H. Liu 1} and holds at the full quantum level. 

Canonically normalizing the fields $\pi^A_{r/a}$, it is convenient to write
\begin{equation} F^{AB}(i\omega, k^2) = i\omega \delta ^{AB} -k^2 \cD^{AB}(i \omega,k^2) + \cR^{AB}(i\omega,k^2).\end{equation}
Suppose we are interested in frequencies and wavelengths much less than $\beta_0$. Then, we can Taylor expand $F^{AB}$ in small $\omega$ and $k^2$. We have at leading order in $\omega$ and $k^2$,
\begin{equation} F^{AB}(i\omega, k^2) = i\omega \delta ^{AB} -k^2 D^{AB} + R^{AB},\end{equation}
for constant matrices $D$ and $R$. Then, ignoring the stochastic field $\zeta^A$ and performing an inverse Fourier transform, the equations of motion become
\begin{equation}  \partial_t \pi^A = D^{AB} \p^2 \pi^B + R^{AB} \pi^B,   \end{equation}
which is the linearized version of the usual reaction-diffusion equation with diffusion matrix $D$ and reaction matrix $R$. Notice also that since $F_{AB}\cP^A_C$ vanishes for $\omega, \vec k\to0$, we have $R^{AB}\cP^A_C=0$, meaning that the hydrodynamical fields corresponding to the exactly conserved $U(1)$ currents are given by the ordinary diffusion equation, with no reaction matrix. 

Finally, excitations that persist over long times correspond to $\omega\to0$, that is
\begin{equation} \squ{k^2 D^{AB}  - R^{AB}}\pi^B=0.\end{equation}
Assuming that $D$ is invertible, the values of $k$ that yield $\omega=0$ are given by 
\begin{equation}0=\det\cur{D^{-1}\cdot R-k^2}.\end{equation}
That is, excitations last for arbitrarily long times as $k^2$ approaches an eigenvalue of $D^{-1}\cdot R$.

\section{Conclusions}\label{Conclusions}

We have developed two formalisms for describing the dynamics of chemical reactions taking place in fluid phase. The first method took the conservation equations of the stress-energy tensor and the $\text{(non-)}$conservation equations of particle-number currents expressed in terms of constitutive relations as the starting point. Then principles from thermodynamics were used to place further constrains on the equations of motion. The advantage of this approach is that the equations of motion follow from relatively straight-forward arguments and it is therefore accessible to most physicists. The second method relied solely on symmetry principles and used them to construct a non-equilibrium effective action. This method, despite using vastly more complicated machinery than the first, has the advantage that everything is derivable from symmetry considerations alone. In particular, thermodynamics was automatically taken into account, so no additional constraints had to be imposed by hand. As a result, extending the derivative expansion to higher orders is a straightforward, almost mechanical procedure. 

Our constructions of reactive hydrodynamics in \S\ref{EOM} and \S\ref{DC EFT} were purely classical. Using the equations of motion approach, extending to the quantum regime would be a highly non-trivial task. However from the perspective of the effective action, it is possible to construct $I_\text{EFT}$ as an expansion in $\hbar$, thereby taking into account  quantum effects perturbatively. Thus the EFT approach is significantly more powerful than the $\text{(non-)}$conservation approach. Further, even the classical, leading-order effective action that we construct contains more information than just the $\text{(non-)}$conservation equations and the entropy current. The terms that contain more than one factor of $a$-type fields correspond to thermal noise~\cite{H. Liu 1}. Thus, the effective action formalism contains information about statistical fluctuations, something that the methods of \S\ref{EOM} know nothing about. Finally, to underscore the power the non-equilibrium EFT approach, we constructed a quadratic action describing the reactive and diffusive dynamics that is valid to all orders in the derivative expansion. This EFT describes the full quantum and thermal fluctuations of the hydrodynamic modes describing reactions and diffusion at the linearized level. Thus, even though the effective action approach requires a lot more machinery, it allows for a much more systematic and complete description of the hydrodynamics of chemical reactions.

The EFT approach to reactive flows has a wide range of possible applications. It can be used to describe condensed matter systems involving chemical reactions as well as systems involving nuclear reactions. Such systems could arise in the lab, in engineering settings, or in astrophysical systems like stars. Additionally, large-scale cosmology is often modeled by fluids that undergo reactions that convert various types of matter and radiation into one another. We therefore expect that our EFT approach could be a valuable tool to cosmologists. 

In addition to having practical benefits, our effective action approach is of some theoretical interest as well. In particular, it applies the machinery of \cite{H. Liu 1,H. Liu 2,H. Liu 3} to systems that exhibit dissipative behavior even in the leading-order hydrodynamical limit. Thus, we are able to use an action to understand a system that, until recently, could not have admitted an EFT description. 

\bigskip

\noindent {\bf Acknowledgments:} I would like to thank Lam Hui and Alberto Nicolis for their wonderful mentorship. This work was partially supported by the US Department of Energy grant $\text{DE-SC0011941}$.

\appendix

\section{The meaning of  \textbf{\emph{r}}-and  \textbf{\emph{a}}-type fields}\label{a-type variables appendix}

In this section, we will investigate the physical meanings of the $r$-and $a$-type fields. In particular, we will see that $r$-type fields play the role of classical field configurations, whereas $a$-type fields encode information about noise due to thermal and quantum fluctuations.  Suppose that we have a non-equilibrium effective action $I_\text{EFT}[\varphi_r,\varphi_a]$. Notice that the third unitarity constraint of (\ref{unitarity constraints}) requires that $I_\text{EFT}$ vanish if we set $\varphi_a=0$, meaning that every term in the effective action must contain at least one $a$-type field. It turns out that the terms in the effective action that are linear in $a$-type fields play very different roles from those with higher powers of $a$-type fields. Therefore, let us write 
\begin{equation}I_\text{EFT}[\varphi_r,\varphi_a] = \int d^4 x~ E[\varphi_r]\varphi_a +I_2[\varphi_r,\varphi_a], \end{equation}
where $I_2$ contains terms that are at least quadratic $\varphi_a$. Next, define $\eta[\varphi_r,\zeta]$ by
\begin{equation}e^{-\eta[\varphi_r,\zeta]}\equiv \int \cD[\varphi_a]\exp\cur{i \int d^4 xE[\varphi_r]\varphi_a +i\int d^4 x \zeta \varphi_a}.  \end{equation}
Using this definition, we have 
\begin{equation}\begin{split} \label{path integral ra}& \int  \cD[\varphi_r\varphi_a]~ e^{i I_\text{EFT}[\varphi_r,\varphi_a]} \\ & =  \int \cD [\varphi_r \varphi_a\zeta] ~ \exp\cur{i\int d^4 x ~\cur{E[\varphi_r]-\zeta} \varphi_a - \eta[\varphi_r,\zeta] } \\
& = \int \cD [\varphi_r\zeta] ~\delta_D[E[\varphi_r]-\zeta] e^{-\eta[\varphi_r,\zeta]},
\end{split} \end{equation}
where $\delta_D$ is the Dirac delta functional. For any given $\zeta$, the argument of the delta functional has a straightforward interpretation; it contains the equations of motion, namely
\begin{equation} E[\varphi_r] = \zeta. \end{equation}
But we see from the r.h.s. of (\ref{path integral ra}) that for each $\zeta$, the delta functional is weighted by $e^{-\eta[\varphi_r,\zeta]}$. We therefore interpret $\zeta$ as a stochastic field with probability distribution $e^{-\eta[\varphi_r,\zeta]}$. 

Using this interpretation of $r$-and $a$-fields, if we take our effective Lagrangian to be (\ref{Fourier Lagrangian}), then we find that the equations of motion are given by (\ref{Fourier eom}), where $\zeta^A$ is a Gaussian-random stochastic variable with two-point function $\vev{\zeta^A\zeta^B}= M^{AB}$.


\end{document}